\documentclass[acmsmall,screen]{acmart}
\usepackage{graphicx}
\usepackage{comment} 
\usepackage{subcaption}
\usepackage{tcolorbox}
\usepackage{tabularx}
\usepackage{booktabs}
\usepackage{multirow}
\usepackage{makecell}
\AtBeginDocument{%
  }

\setcopyright{none}
\renewcommand\footnotetextcopyrightpermission[1]{}

\begin{document}
\title{Can LLMs Fix It Without Code? Toward Automated Verification of No-Code Bug Fixes
}

\author{Utku Boran Torun}
\authornote{Both authors contributed equally to this research.}

\email{boran.torun@bilkent.edu.tr}
\orcid{0009-0000-9571-6287}
\affiliation{%
  \institution{Bilkent University}
  \city{Ankara}
  \country{Turkey}}

\author{Veli Karakaya}
\authornotemark[1]
\email{veli.karakaya@ug.bilkent.edu.tr}
\orcid{0009-0009-5389-6588}
\affiliation{%
  \institution{Bilkent University}
  \city{Ankara}
  \country{Turkey}}

\author{Eray Tüzün}
\email{eraytuzun@cs.bilkent.edu.tr}
\orcid{0000-0002-5550-7816}
\affiliation{%
  \institution{Bilkent University}
  \city{Ankara}
  \country{Turkey}
}

\renewcommand{\shortauthors}{}

\begin{abstract}
\noindent\textbf{Context.} A \emph{no-code fix} resolves an invalid bug report by directing the user to change a setting, update to a version where the problem is already fixed, or adjust their workflow, rather than by modifying the source code. Bug reports frequently admit such fixes, but manually verifying whether a proposed no-code fix genuinely resolves the reported bug consumes considerable developer time that could otherwise go toward core development work.

\noindent\textbf{Objective.} This study proposes and evaluates an automated pipeline for the capability of large language models (LLMs) to generate no-code fixes, replacing manual verification with execution-based evaluation in a real browser environment.

\noindent\textbf{Method.} We evaluate 322 no-code fixes generated by the 12 LLM-based configurations released with the benchmark of a previous study. From this benchmark, we select the bug reports categorized as Faulty Configuration, Wrong Version, or External System \& Dependency. For each report, evaluation proceeds through an execution pipeline in which an executor agent applies each fix by following its natural-language instructions in a real browser environment, and an issue-specific checker determines whether the reported bug persists. We repeat the full pipeline with three executors from different providers: two Computer-Use Agents (CUAs), OpenCUA-72B and Anthropic's Claude Sonnet~5 with computer use, and one multimodal agentic LLM, Meta's Muse Glimmer, to characterize how much the choice of executor affects the outcome.

\noindent\textbf{Results.} 17.6\% of the candidate issues could be set up in the environment and passed both sanity gates, and the remaining issues could not be reproduced. Across the 322 fixes 14.6\% to 49.7\% resolved the bug depending on the executor, and the strongest configuration, Claude Opus~4.6 in the Vanilla pipeline, resolved up to 74.1\% of its fixes under the executor Claude Sonnet~5. Changing only the executor shifted a configuration's resolution rate by 38.8\% on average, and the three executors reached the same verdict on only 46.9\% of the fixes. Compared with human execution, the executors matched the outcome on which two human executors agreed for 66.1\% to 88.1\% of the sampled fixes.

\noindent\textbf{Conclusion.} Even under the best executor, fewer than half of the LLM-generated no-code fixes resolve the reported bug, so such fixes need verification before they reach users. Execution-based verification can provide this without manual inspection, but the measured capability depends strongly on the executor, which evaluations must therefore report and control.
\end{abstract}




\maketitle
\pagestyle{plain}
\thispagestyle{plain}

\section{Introduction}
\label{sec:intro}
Large language models (LLMs) have rapidly become part of everyday software
engineering practice. They now assist developers across the software development lifecycle,
powering code generation \cite{jiang2026survey, zhang2023planning, jiang2024self}, automated code review \cite{pornprasit2024fine, cihan2025automated, yu2024fine}, test generation \cite{schafer2023empirical, celik2025review, zhang2024testbench}, and bug
triage \cite{du2024llm, aracena2024applying, colavito2025benchmarking}. As these systems move from research prototypes into tools
developers rely on daily \cite{stackoverflow2025survey} software engineering tasks, they are increasingly applied not only to
writing and analyzing code, but to the surrounding work of managing
the issues that software projects accumulate \cite{jimenez2024swe}.

One of the most time-consuming parts of that work is triaging bug
reports, and in particular dealing with reports that turn out not to
require a code change at all \cite{gon2026automatedrootcausesubclassificationnocode}. A bug report is \emph{invalid} when it does not describe a defect in the
source code: the reported behavior instead stems from user
misconfiguration, an outdated version, a conflicting extension, or a
misunderstanding of intended behavior, or the report is not considered as a bug report
at all, such as a feature request or a question. Such reports make up a
substantial share of what projects receive~\cite{fan2018chaff,
herzig2013its}: a manual inspection of more than 7{,}000 reports across
five open-source projects found that 33.8\% of those filed as bugs were
misclassified~\cite{herzig2013its}. Rather than with a patch, many invalid
reports are resolved by directing the reporter to adjust a setting, update
to a version where the problem is already fixed, or disable a conflicting
extension. Following Gön et
al.~\cite{gon2026automatedrootcausesubclassificationnocode}, we call such a
resolution a \emph{no-code fix}. Triage already consumes a large fraction of maintenance effort, and
invalid reports add to that burden without leading to a code change: in
an industrial case study, Laiq et al.~\cite{laiq2022early} found that 14
to 17\% of submitted reports were invalid, yet these took roughly 16 to
18 calendar days to resolve on average, about as long as valid reports
despite requiring no code change. A maintainer must still read the
report, reproduce or investigate the problem, determine that no code fix
is needed, and explain the resolution to the
reporter~\cite{xuan2014towards,
gon2026automatedrootcausesubclassificationnocode}.

Gön et al.~\cite{gon2026automatedrootcausesubclassificationnocode} introduced a
benchmark of invalid Brave\footnote{https://brave.com/} browser bug reports,
subclassify them by root cause, and generate no-code fixes for them with
several LLM configurations, the same benchmark and fixes we evaluate in this
paper. Evaluating such fixes is hard. The only reference is the maintainer's
resolution, a free-form comment written for one reporter in one situation, so
two fixes can be worded differently while doing the same thing, or worded
almost identically while one points to a setting that does not exist. Their study therefore relies on text similarity and an LLM judge,
the usual fallbacks when an output cannot be checked by running
it~\cite{torun2026evaluationllmbasedsoftwareengineering}. Their
BERTScore~\cite{zhang2019bertscore} values fall within a narrow band of 0.81 to
0.82 across all models and configurations, and the LLM judge, while more
discriminative, inherits the documented biases of LLM
judges~\cite{shi2025judging}. Neither shows whether a fix would work if a user
followed it (Section~\ref{sec:discussion-bertscore} revisits this on the fixes
we execute). This is the gap we address.
Execution offers an alternative. A code patch either passes the test suite or
not, and Zhang et al.~\cite{zhang2026code} have extended this principle to code
review by turning review comments into executable tests. A no-code fix produces
no code change, but its instructions can still be carried out: changing a
setting or disabling an extension is an action in a real environment, after
which the reported symptom either persists or disappears. Our goal is
therefore to evaluate LLM-generated no-code fixes by whether they actually resolve the
reported problem, not by whether they resemble a fix that did.
Recent Computer-Use Agents (CUAs) make this process practical to automate, since they
interpret natural-language instructions and act on a real interface, which is
exactly what applying a no-code fix requires.

Building on this observation, we design an automated, execution-based
pipeline for evaluating no-code fixes. For each bug report, the pipeline first applies the maintainer's fix and
then rolls it back, which yields a verified fixed state and, from it, a
verified buggy state; it then applies a candidate fix to the buggy state
and checks whether the bug persists afterwards. Every candidate fix is applied by an executor agent directly from its natural-language text and verified by a deterministic, issue-specific checker; the bug-specific parts of the pipeline are confined to setting up each issue's environment and its checker. We run it with three executors (Claude Sonnet 5 \cite{anthropic2026sonnet5}, OpenCUA-72B \cite{wang2025opencuaopenfoundationscomputeruse}, Meta Muse Glimmer \cite{meta2026museglimmer}) from different
providers and validate the automated verdicts against humans performing
the same procedure by hand. Concretely, we investigate three research questions:

\begin{itemize}
    \item \textbf{RQ1: How effectively do agentic executors resolve LLM-generated no-code fixes?} We measure the fix-generation
    capability of 12 LLM-based configurations by executing their
    fixes end-to-end, reporting resolution and applicability rates
    overall and broken down by bug subclass.

    \item \textbf{RQ2: To what extent does the choice of executor affect the measured fix capability?} We run the same fixes
    through three executors from different providers to separate variance
    introduced by the executor from variance in the underlying fix
    quality, and to test whether conclusions about LLM performance are
    stable across executor choice.

    \item \textbf{RQ3: How accurately do agentic executors execute LLM-generated
    fixes compared to human execution?} On a sampled subset of fixes,
    we compare each executor's outcome against a human performing the same
    procedure by hand, establishing how far the automated verdicts can
    be trusted.
\end{itemize}

This paper makes the following contributions:

\begin{itemize}
    \item We introduce a rollback-based, execution-driven pipeline for
    evaluating LLM-generated no-code fixes. Two sanity gates confirm that
    the bug is absent once the maintainer's fix is applied and present once
    it is rolled back, so every candidate fix runs on a verified buggy state
    and is judged by whether it removes the bug.

    \item We instantiate the pipeline with agentic executors that apply
    each natural-language fix directly, verified by 
    issue-specific state check, and we compare three executors, two CUAs
    and one multimodal agentic LLM, to quantify the effect of executor
    choice.

    \item We evaluate 322 individual no-code fixes (12 LLM-based configurations $\times$ three generation runs $\times$ nine reproducible bug reports, of which two generations are missing from the source benchmark), reporting per-configuration resolution and applicability rates at the level of individual fixes and their variation across bug subclasses.

    \item We compare the automated verdicts with human execution of the
    same fixes, indicating how far each executor's verdicts can be
    trusted and in which direction they diverge.
\end{itemize}

\section{Related Work}
\label{sec:related}
\subsection{Invalid Bug Reports: Classification and No-Code Fix Generation}

Not all issues submitted to a bug tracker call for a code
change. Herzig et al.~\cite{herzig2013its} manually examined more
than 7,000 issue reports across five open-source projects and
found that 33.8\% were misclassified: they did not point to a
defect but requested a new feature, a documentation
update, or an internal refactoring, introducing measurable bias
into downstream bug-prediction studies. Distinguishing such
reports from genuine defects has since become a task in its
own right, studied extensively before the arrival of LLMs.
Antoniol et al.~\cite{antoniol2008bug} and Zanetti et
al.~\cite{zanetti2013categorizing} used text-based and social-network
features, respectively, to separate genuine bugs from other change
requests; Fan et al.~\cite{fan2018chaff}, Otoom et
al.~\cite{otoom2019automated}, and Pandey et
al.~\cite{pandey2017automated} trained SVM and random-forest
classifiers for the same task, with Fan et al. identifying textual
content and submitter experience as the strongest predictors.
Meng and Visser~\cite{meng2023bug} fine-tuned BERT to classify
and explain bug report validity, and Laiq et
al.~\cite{laiq2025comparative} compared classical and
transformer-based models on a unified benchmark, finding
fine-tuned RoBERTa to perform best. Du et al.~\cite{du2024llm}
propose LLM-BRC, a large-language-model--based framework for
bug report classification; Aracena et al.~\cite{aracena2024applying}
and Colavito et al.~\cite{colavito2025benchmarking} likewise apply
and benchmark LLMs for automated issue labeling. Closer to our
setting, Din{\c{c}} and T{\"u}z{\"u}n~\cite{dincc2025judge} apply a
range of LLM and machine-learning methods specifically to the
valid/invalid distinction, characterizing where LLMs succeed and
fail at this classification.

Once a report is identified as invalid, resolving it typically requires
no patch, only an explanation of a configuration change, a version update,
or a workaround. This is what we call a no-code fix. G\"on
et al.~\cite{gon2026automatedrootcausesubclassificationnocode} introduce
a benchmark of invalid Brave browser issues, subclassify them by root
cause, and generate no-code fixes for them using 12 LLM-based
configurations (Section~\ref{sec:llm-configs}). We build directly on this
benchmark (Section~\ref{sec:dataset}). Torun et al.~\cite{torun2026past} situate this task within a broader vision for artificial intelligence (AI)-assisted bug tracking, in which incoming reports are automatically triaged as valid or invalid, invalid reports are resolved by agent-suggested no-code fixes, and valid ones proceed to patch generation and human review. Our work can be read as an empirical test of whether the no-code-fix step of that vision is achievable with current LLMs. Where prior work \cite{gon2026automatedrootcausesubclassificationnocode} in this space evaluates generated fixes by comparing them to a reference resolution, we evaluate by execution: applying each fix to a live environment and checking whether the reported bug persists.

\subsection{Agent-Based Bug Reproduction and Fix Verification}

A separate line of work uses LLMs and agents to interact with running software. Kang et al.~\cite{kang2024evaluating} prompt LLMs to generate bug-reproducing tests directly from natural-language bug reports and evaluate 15 models on the Defects4J benchmark, reproducing roughly one third of the included bugs; their technique, like later agentic approaches~\cite{cheng2025agentic,guo2026swe}, operates at the level of unit tests rather than a live GUI. Feng and Chen~\cite{feng2024prompting} instead target a live GUI directly: their AdbGPT system prompts an LLM to extract steps-to-reproduce from an Android bug report and replays them as GUI events on the device without any bug-specific scripting, closer in spirit to how our pipeline executes steps-to-reproduce in a browser. Closer to our environment, Yapa\u{g}c{\i} et al.~\cite{yapaugci2025agents} introduce BugCraft, which uses an LLM to synthesize steps-to-reproduce from a Minecraft crash report and a vision-based agent to execute them in the game, reproducing 34.9\% of crash bugs end-to-end and outperforming baseline computer-use agents by 37\%; our sanity checks follow the same execute-steps-and-observe-the-outcome principle, applied to a browser instead of a game. Chen et al.~\cite{chen2025standing} take a complementary retrieval-augmented approach, using bug reports from similar Android apps to guide GUI exploration toward paths likely to trigger a bug, again treating a bug report as an executable specification for an agent.

The work closest to ours is Fixpad++~\cite{ir2026fixpad++}, which verifies whether a candidate patch fixes a reported crash in the GUI-based Notepad++ editor: a multi-agent system reproduces the bug via an extended ReAct loop with self-reflection, then replays the same trajectory on the patched build to judge fix correctness. Fixpad++ and our pipeline share the core idea of verifying a fix by re-executing a reproduction trajectory rather than by comparing text, but differ in what is being verified and how: Fixpad++ judges source-code patches with a purpose-built multi-agent
architecture tuned to a single application. We instead judge
natural-language no-code fixes with off-the-shelf agentic executors, both
open-weight and proprietary, and treat the choice of executor itself as an
independent variable (RQ2).

Agent-based verification of GUI artifacts has also been studied outside the bug-fixing setting. Kolthoff et al.~\cite{kolthoff2025guispector} introduce GUISpector, which uses a multi-modal LLM agent, built on a commercial CUA, to autonomously verify whether a GUI prototype satisfies natural-language requirements and to report which acceptance criteria are met or violated; unlike Fixpad++ and our pipeline, GUISpector checks conformance to a specification, not whether a previously observed bug has disappeared. Closer to our concern with executor reliability, Hong et al.~\cite{hong2026diageval} show that GUI-agent evaluators frequently misattribute their own execution failures to genuine software defects, and propose reusing a failed trajectory to disambiguate the two; this distinction between evaluator error and actual software defect is precisely what our sanity checks and our RQ3 human-comparison are designed to control for in our own pipeline.

Taken together, this line of work reproduces bugs in unit-test, game, and mobile-GUI settings~\cite{kang2024evaluating,cheng2025agentic,guo2026swe,yapaugci2025agents,chen2025standing,feng2024prompting}, verifies fixes with a bespoke agent tied to a single desktop application~\cite{ir2026fixpad++}, or verifies specification conformance rather than bug resolution~\cite{kolthoff2025guispector}. None evaluates no-code fixes specifically, and none treats the choice of executing agent as an experimental variable, which is the focus of our RQ2.

\subsection{LLM Evaluation in Software Engineering}

Evaluating LLM-based software engineering tools reliably is difficult in its own right. Torun et al.~\cite{torun2026evaluationllmbasedsoftwareengineering} survey current evaluation practices and argue that the open-endedness and non-determinism of LLM outputs undermine the single-ground-truth, deterministic assumptions that conventional SE evaluation relies on. One increasingly common workaround, the LLM-as-a-Judge paradigm, used by G\"on et al.~\cite{gon2026automatedrootcausesubclassificationnocode} alongside BERTScore to grade generated no-code fixes, is itself the subject of active scrutiny: He et al.~\cite{he2025code} survey LLM-as-a-Judge practice in software engineering and identify a lack of validation against ground truth as one of the paradigm's central open problems.

Unlike execution-free critics~\cite{yadavally2025large}, execution-based evaluation sidesteps this problem directly, and it has an established track record for LLM-generated code. SWE-bench~\cite{jimenez2024swe} and SWE-bench-Live~\cite{zhang2026swe} evaluate whether a model's patch to a real codebase makes a GitHub issue's associated tests pass, instead of scoring the patch's similarity to a reference diff. The same execution-based principle has recently been extended from source code to the GUI: OSWorld~\cite{xie2024osworld} scores agents on 369 real desktop and web tasks by checking the resulting operating-system state against a task-specific verification script, and WebArena~\cite{zhou2024webarena} evaluates web agents on self-hosted, fully functional websites by checking the functional correctness of the completed task. Our pipeline borrows this state-based checking but inverts the role of the agent: in OSWorld and WebArena the agent is the subject under evaluation, whereas here the agentic executor is part of the measurement instrument and the subject is the LLM-generated no-code fix it applies. This is why we treat the executor as an experimental variable (RQ2) and validate it against human execution (RQ3).

\section{Methodology}
\label{sec:methodology}

This section describes the dataset we work
with, the rollback-based evaluation pipeline shown in Figure \ref{fig:pipeline-overview}, the use
of three agentic executors as pipeline executors, and the metrics we
report.

\subsection{Dataset}
\label{sec:dataset}

We reuse the benchmark released by Gön et
al.~\cite{gon2026automatedrootcausesubclassificationnocode}, which
contains 302 annotated bug reports drawn from the Brave browser GitHub
issue tracker. Each report is annotated with an invalid-bug-report
subclass and, where available, a link to the maintainer comment
containing the successful no-code fix. Of the 302 reports, 188 fall
into one of seven invalid subclasses, while the remaining 114 are
outliers that the original annotation placed outside the taxonomy
(reports found to be valid, internal developer-workflow items, or cases
with too little information to classify). The benchmark also includes no-code fixes
generated by 12 LLM-based configurations for each bug report, spanning
four pipeline variants and six backbone LLMs; we describe these
configurations in detail in Section~\ref{sec:llm-configs}. Each
configuration was executed three times in the original study
to account for LLM stochasticity, yielding 36 candidate no-code fixes
per bug report (12 configurations $\times$ three runs), all of which we
evaluate independently in our pipeline.

Not every subclass in the benchmark is suitable for execution-based
evaluation. Fixes for the \textit{Working as Designed} (48 reports),
\textit{Feature Request} (41), \textit{Non-Reproducible} (35), and
\textit{Question} (13) subclasses do not correspond to actions that can
be applied and verified in a browser environment: they are
explanations, requests for information, or acknowledgments none of which
changes the application state. We therefore select the 51 bug reports in the three
subclasses whose maintainer fix takes the form of a concrete action on
the browser or its environment: \textit{Faulty Configuration \&
Environment Setup} (19), \textit{Wrong Version (Already Fixed)} (12),
and \textit{External System \& Dependency Issues} (20).

\subsection{LLM-Based Fix-Generation Configurations}
\label{sec:llm-configs}

For each bug report, the benchmark provides no-code fixes
generated by 12 distinct configurations, spanning four pipeline
variants and six backbone LLMs. Table~\ref{tab:llm-configs}
lists all 12 configurations together with the short identifiers we
use to refer to them for the remainder of the paper.

\textbf{Vanilla.} The baseline pipeline prompts a single backbone LLM
with the bug report and its predicted invalid-subclass label, and is
instantiated with six backbone models: three proprietary (Gemini~3.1 
Pro Preview \cite{google2026gemini31pro}, GPT-5.4 \cite{openai2026gpt54}, and Claude Opus~4.6 \cite{anthropic2026opus46}) and three open-weight
(GLM-5 \cite{zeng2026glm}, MiniMax~M2.7 \cite{minimax2026m27}, and Kimi~K2.5 \cite{team2026kimi}).

\textbf{Ablation (No-Prior).} Identical to Vanilla but withholds the
invalid-subclass prior from the prompt, isolating its contribution to
fix quality. Instantiated with the best-performing proprietary and
open-weight backbones from the Vanilla setting (Gemini~3.1 Pro and
Kimi~K2.5).

\textbf{RAG.} Retrieves from two corpora, Brave Browser wiki pages and
historical invalid bug reports, embedded with the Zembed-1 model\footnote{\url{https://huggingface.co/zeroentropy/zembed-1}} and
indexed in LanceDB\footnote{\url{https://www.lancedb.com/}}. For each query issue, the pipeline retrieves the
top-20 nearest neighbors from each corpus via dense vector search,
discards candidates created after the query issue (chronological
filtering, to prevent data leakage), and re-ranks the remaining
candidates with the Zerank-2\footnote{\url{https://huggingface.co/zeroentropy/zerank-2}} cross-encoder to select the top-5
contexts, which are appended to the prompt. Instantiated with the same
two backbones as the ablation setting (Gemini~3.1 Pro and Kimi~K2.5).

\textbf{Agentic Web Search.} Follows a WebThinker-style
``Think-Search-and-Draft'' architecture: the backbone LLM first drafts
a research brief, then a bounded orchestration loop issues queries against the Serper Search
API. The canonical original-issue page is excluded from results to
prevent leakage, and the content of any visited page is compressed
before being returned. Retrieved evidence is passed back to the LLM in
chronological order before it produces the final classification and
fix. Instantiated with Gemini~3.1 Pro Preview (paired with
Gemini~2.5~Flash for the search loop) and Kimi~K2.5.

\begin{table}
  \centering
  \footnotesize
  \setlength{\tabcolsep}{4pt}
  \caption{The 12 LLM-based fix-generation configurations reused
  from Gön et al.~\cite{gon2026automatedrootcausesubclassificationnocode},
  with the short identifiers used throughout this paper. Rows are
  pipelines and columns are backbone LLMs; a dash marks a combination
  that was not run.}
  \label{tab:llm-configs}
  \begin{tabular}{@{}l cccccc@{}}
    \toprule
    & \multicolumn{6}{c}{\textbf{Backbone LLM}} \\
    \cmidrule(l){2-7}
    \textbf{Pipeline}
      & \makecell{\textbf{Gemini 3.1}\\\textbf{Pro Preview}}
      & \textbf{GPT-5.4}
      & \makecell{\textbf{Claude}\\\textbf{Opus 4.6}}
      & \textbf{GLM-5}
      & \makecell{\textbf{MiniMax}\\\textbf{M2.7}}
      & \textbf{Kimi K2.5} \\
    \midrule
    Vanilla            & V-Gemini       & V-GPT & V-Claude & V-GLM & V-MiniMax & V-Kimi \\
    \cmidrule{1-7}
    \makecell[l]{Ablation\\(no subclass prior)}
                       & Abl-Gemini     & --    & --       & --    & --        & Abl-Kimi \\
    \cmidrule{1-7}
    RAG                & RAG-Gemini     & --    & --       & --    & --        & RAG-Kimi \\
    \cmidrule{1-7}
    Agentic web search & Agentic-Gemini & --    & --       & --    & --        & Agentic-Kimi \\
    \bottomrule
  \end{tabular}
\end{table}

\begin{figure*}
    \centering
    \includegraphics[width=\linewidth]{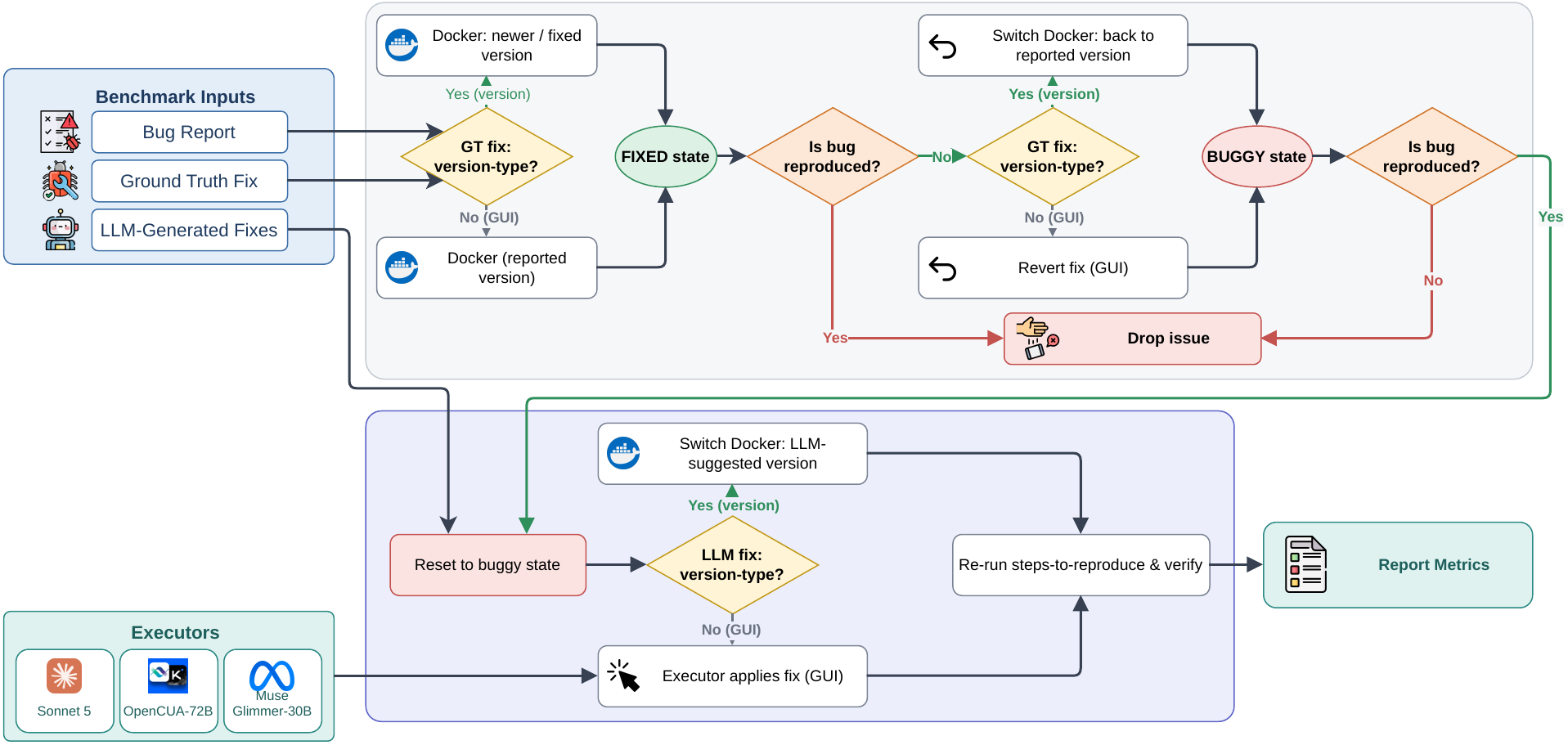}
    \caption{Overview of the evaluation pipeline.}
    \Description{Flow diagram: a bug report, the maintainer's fix and 12 LLM fixes enter a Docker environment; two sanity checks bracket a rollback from fixed to buggy state; each LLM fix is applied by an executor and checked.}
    \label{fig:pipeline-overview}
\end{figure*}

\subsection{Pipeline Overview}
\label{sec:pipeline}

For each of the 51 selected bug reports, evaluation proceeds through the
pipeline shown in Figure~\ref{fig:pipeline-overview}. The pipeline takes
three inputs on the left: the bug report itself (including its
steps-to-reproduce and stated actual result), the maintainer's
ground-truth fix, and the 12 candidate LLM-generated fixes. It produces a per-LLM resolution rate at the right, indicating how often each LLM's fix resolved the reported bug.

The main flow is bracketed by two sanity checks and separated by an
explicit state rollback. A further branch, based on whether the ground-truth fix is a version update or a GUI-level change, determines how that fix is applied and rolled back. We describe each stage in turn, following
Figure~\ref{fig:pipeline-overview} left to right.

\paragraph{Fresh environment and ground-truth application.}
For each bug report, we classify the maintainer's ground-truth fix as
\emph{version-type} (the bug is resolved simply by moving to a newer
Brave release) or \emph{GUI-type} (the bug is resolved by a
configuration change, extension toggle, or other in-browser action).
This classification follows the content of the maintainer's resolution
(\emph{GT fix: version-type?} in Figure~\ref{fig:pipeline-overview}). The
Wrong Version issue is version-type by definition, and so are two external
system issues whose maintainer resolved the report by pointing to a later
release that contains an upstream Chromium fix; the remaining six issues
are GUI-type. For a version-type fix, we
provision a Docker container already running the newer, fixed Brave
release (\emph{Docker: newer / fixed version}). For a GUI-type fix, we
provision a container running the version originally reported by the
issue reporter (\emph{Docker (reported version)}), and we apply
the maintainer's fix to it through the browser GUI. Either path yields
the \emph{fixed} state (green ellipse in Figure~\ref{fig:pipeline-overview}).

\paragraph{Fixed-state sanity check.}
Before rolling anything back, we verify that the fixed state is
genuinely fixed: the steps-to-reproduce from the bug report are
executed and checked against the report's stated actual result
(\emph{Is bug reproduced?} in Figure~\ref{fig:pipeline-overview}). If the
actual result is \emph{not} observed, this check passes and the
pipeline continues to the rollback stage. If it \emph{is} observed, the
ground-truth fix as applied does not resolve the bug in our
environment; the issue is dropped (\emph{Drop issue}) and no further
evaluation is attempted for it.

\paragraph{Rollback to the buggy state.}
Once this check passes, the ground-truth fix is rolled back, branching
again on whether it was version-type or GUI-type. For a version-type
fix, we switch the Docker container back to the originally reported
version (\emph{Switch Docker: back to reported version}). For a
GUI-type fix, a scripted sequence of GUI actions reverts the fix
without changing the container version (\emph{Revert fix (GUI)}).
Either path returns the environment to the configuration that
exhibited the bug before the fix, which we call the \emph{buggy} state
(red ellipse in Figure~\ref{fig:pipeline-overview}). The rollback
confirms that the state the checker reads in the buggy environment
corresponds to the reported symptom.

\paragraph{Buggy-state sanity check.}
We now verify that the rollback actually produced the reported bug:
the steps-to-reproduce are executed again and checked against the
stated actual result (the second \emph{Is bug reproduced?} check in
Figure~\ref{fig:pipeline-overview}). If the actual result \emph{is}
observed, this check passes, and the issue proceeds to per-LLM
evaluation. If not, the rollback failed to reproduce the bug, and the
issue is dropped (\emph{Drop issue}). Together, these two checks
ensure that every issue carried into the LLM evaluation stage has a
genuinely fixed reference state and a genuinely buggy state derived
from it. Of the 51 selected issues, 38 could not be set up in our
environment at all, for example because they were specific to a
platform other than Linux or required a live backend service, and four
did not exhibit the reported bug when tested; the remaining nine passed
both checks and formed the effective evaluation set.

\paragraph{Per-LLM fix application and verification.}
For each of the 36 candidate fixes (12 configurations, three
independently generated runs each), every episode starts from a fresh
container in the buggy state verified above (\emph{Reset to buggy
state}): for GUI-type issues, a fresh profile on the gate-verified
build, with the buggy value pre-seeded where the bug is not Brave's
default state; for the version-type
issues, the originally reported build. The executor then applies the
LLM's fix through the browser GUI (\emph{Executor applies fix (GUI)}).
On the version-type issues, a fix may instead call for an update: when
the executor opens Brave's update page (\texttt{brave://settings/help}),
the harness switches the container to the fixed build (\emph{Switch
Docker} in Figure~\ref{fig:pipeline-overview}), which is what a user
would obtain by updating. Because the switch is triggered by the
executor's own action, resolution rates on these issues can differ
across executors. After the episode, an issue-specific checker reads
the resulting browser state and decides whether the reported actual
result is still present; we additionally replay the steps-to-reproduce
and keep the screenshots as a visual record (\emph{Re-run
steps-to-reproduce \& verify}). This verdict is produced by the checker (Section~\ref{sec:cuas}). Each attempt
therefore yields two independent signals: the checker's verdict on
whether the bug persists, and the executor's own termination status
(DONE, BLOCKED, step budget exhausted, or stopped without a verdict),
which feed the metrics in Section~\ref{sec:metrics}.

\subsection{Executors and Checkers}
\label{sec:cuas}

Executors are used in the per-fix stage: every candidate fix is applied
by an agentic executor instead of a bug-specific script, which is
what allows the pipeline to consume the natural-language fixes
directly. The sanity gates and the rollback
(Section~\ref{sec:pipeline}) are scripted, so that the validity of each
issue's environment does not depend on the agents under study.

To isolate variance introduced by the choice of executor from variance
in the underlying LLM fix quality, we run the per-fix stage once with
each of three executors. The three executors we compare are OpenCUA-72B~\cite{wang2025opencuaopenfoundationscomputeruse}, an
open-weight model trained for computer use; Meta's Muse
Glimmer~\cite{meta2026museglimmer}, an open-weight general-purpose
agentic model; and Anthropic's Claude Sonnet~5 \cite{anthropic2026sonnet5} with its computer-use
tool (bottom left side of Figure~\ref{fig:pipeline-overview}). All three act on
the same browser environment through a shared action interface. OpenCUA
and Sonnet~5 use the action formats defined by their developers, while
Muse Glimmer, which has no official computer-use harness, is driven
through a function-calling schema we designed; implementation details
are provided in our replication package (Section~\ref{sec:data}). All
three executors receive the same task prompt, containing the bug report and
the fix text, and act for at most 20 steps per episode. Sonnet~5 uses its
default sampling settings, OpenCUA-72B greedy decoding (temperature 0) as in
its reference evaluation code, and Muse Glimmer temperature 1.0 and top-$p$ 0.95 as recommended by its model card\footnote{\url{https://huggingface.co/meta-models/Muse-Glimmer-30B}}.

\paragraph{Checker.} The executor applies a fix but does not decide whether
the fix worked. For each issue, we implement a deterministic checker that
reads the browser state tied to the reported symptom, such as a value in
the browser's preference files or a property of the rendered page queried
through the Chrome DevTools Protocol, and reports whether the reported
actual result is still present. The executor's own claim of success is
recorded but never used as the verdict. The checkers are the only
issue-specific code in the pipeline and are part of our replication
package.

\paragraph{Repetition and aggregation.} Because executor behavior is
stochastic, each fix is executed three times by each executor. We count a fix
as \emph{resolved} under an executor if the checker reports the bug as absent in
at least one of the three runs. The fix text is identical across runs, so if the checker finds the bug
absent after any run, following the fix's instructions can remove the
bug; a run in which the same fix fails may instead reflect an executor
error. We therefore treat one successful run as sufficient, an assumption
that RQ3 tests against human execution. We count a fix as \emph{non-applicable} under an executor if
the executor declared it BLOCKED in at least two of the three runs.

\subsection{Human Execution for Executor Accuracy}
\label{sec:human-eval}

The pipeline's verdicts depend on the executors correctly applying LLM-generated
fixes and correctly observing whether the reported bug persists afterwards.
To measure how reliably they do this, we compare the executors against humans
performing the same task manually.

We draw a simple random sample of 75 of the 322 individual fixes in the evaluation set, using a fixed random seed. This sample size
yields a 95\% confidence level with a margin of error of $\pm 10\%$ for
proportions estimated over the population of 322 fixes, following
Cochran's formula with finite population correction~\cite{cochran1977sampling}
and assuming maximum variability ($p = 0.5$).
For each sampled fix, the buggy state is reproduced identically to how the
executors experience it: a fresh Brave environment with the ground-truth fix
rolled back. A human then executes the LLM's fix on this environment
literally, performing the actions the fix describes without judging their
correctness. The human then runs the steps-to-reproduce and records whether
the reported actual result is observed.

\paragraph{Executor assignment and blinding.}
The human executors are the first two authors. Each author independently
executes all 75 sampled fixes, which allows us to measure inter-rater
agreement between the two executors. Before executing a fix, an author
sees neither the configuration that generated it nor any executor's outcome
for it.

Because the browser state determines the outcome, the comparison does not
rest on subjective agreement. If the executor and
the human executed the same fix on the same starting state, both should in
principle arrive at the same outcome. A divergence may reflect an executor or checker error, an ambiguous fix that admits more than one reasonable reading, or an imperfect human execution, so we treat the human outcome as a reference rather than a ground truth. We
compare each executor's resolved verdict with the
human outcome and report the percentage of matching fixes, with 95\%
Wilson \cite{wilson1927probable} confidence intervals, as the accuracy of that executor.

\subsection{Metrics}
\label{sec:metrics}

We report the following metrics computed over the outcomes produced by the
pipeline:

\textbf{Resolution rate:} For each (configuration, executor) pair, the fraction of the configuration's individual fixes (27 fixes for 11 configurations and 25 for Agentic-Kimi) that the executor resolved, as defined in Section~\ref{sec:cuas}. Aggregated across all 12 configurations, this yields a resolution rate over the 322 fixes per executor. This fix-level rate, rather than an issue-level one, is the primary measure of fix quality in this paper.

\textbf{Applicability rate:} The fraction of fixes that the executor did not declare BLOCKED in the majority of its runs, regardless of whether the bug was ultimately resolved. This separates fixes the executor judged non-actionable (advice, requests for information, deferrals) from applied fixes that failed to resolve the bug. Because it relies on the executor's judgment, applicability is reported per executor.

\textbf{Subclass breakdown:} Fix-level resolution and applicability rates, computed over the relevant subset of the 322 fixes, broken down by the three subclasses in the evaluation set: Faulty Configuration \& Environment Setup, Wrong Version (Already Fixed), and External System \& Dependency Issues.

For RQ2, we additionally report per-LLM resolution rates side by side
across the three executors, and characterize the spread as a measure of
executor sensitivity. For RQ3, we compare each executor's aggregated verdict with the
human-executed outcome on the random sample described in
Section~\ref{sec:human-eval}, reporting the percentage of matching
outcomes with 95\% Wilson intervals, Cohen's $\kappa$ \cite{cohen1960coefficient} on the consensus
set, and the direction of mismatches.

\subsection{Use of generative AI}
We used Claude Sonnet~5~\cite{anthropic2026sonnet5}
to assist in writing the evaluation harness (container orchestration,
executor adapters, and the Muse Glimmer function-calling schema), the
issue-specific checkers, the human-evaluation tool, the analysis
scripts, and the documentation of the replication package. The authors
reviewed this code, verified every checker and scripted gate action
against a live container before use.

\section{Results}
\label{sec:results}

\subsection{RQ1: Fix-Generation Capability of LLMs}
\label{sec:results-rq1}

\paragraph{Overall.}
Table~\ref{tab:rq1-overall} summarizes resolution and applicability per
executor. Across the 322 fixes, the resolution rate was 49.7\% under Sonnet~5,
45.7\% under OpenCUA-72B, and 14.6\% under Muse Glimmer. Applicability was
66.1\%, 99.1\%, and 70.5\%, respectively. To separate fixes that proposed
no executable action from fixes that were applied but did not remove the
bug, we also compute the resolution rate among applicable fixes only,
that is, the share of applicable fixes that resolved the bug. This rate
was 69.5\% under Sonnet~5, 45.8\% under OpenCUA-72B, and 18.1\% under Muse Glimmer.

\begin{table}[h]
  \centering
  \footnotesize
  \caption{Resolution and applicability per executor over the 322 fixes (fix
  counts in parentheses). The last column restricts the resolution rate
  to the fixes the executor found applicable.}
  \label{tab:rq1-overall}
  \begin{tabular}{@{}l r r r@{}}
    \toprule
    \textbf{Executor} & \textbf{Resolution (\%)} & \textbf{Applicability (\%)}
                 & \textbf{Resolution among applicable (\%)} \\
    \midrule
    Claude Sonnet 5 & \textbf{49.7 (160)} & 66.1 (213) & \textbf{69.5 (148)} \\
    OpenCUA-72B     & 45.7 (147) & \textbf{99.1 (319)} & 45.8 (146) \\
    Muse Glimmer    & 14.6 (47)  & 70.5 (227) & 18.1 (41) \\
    \bottomrule
  \end{tabular}
\end{table}

\paragraph{Per configuration.}
Table~\ref{tab:rq1-configs} reports resolution and applicability for each
of the 12 configurations. V-Claude had the highest resolution rate under Sonnet~5 (74.1\%), but not under the other two executors: under OpenCUA-72B, RAG-Kimi and Agentic-Gemini led (55.6\%) and V-Claude was third (51.9\%, tied with V-Gemini), and under Muse Glimmer, Agentic-Kimi led (28.0\%) and V-Claude tied for second (25.9\%). The lowest rate was 25.9\% under Sonnet~5 (V-MiniMax), 33.3\% under OpenCUA-72B (V-Kimi), and 0.0\% under Muse Glimmer (V-Kimi). OpenCUA-72B declared at most one fix per configuration non-applicable, whereas under Sonnet~5 applicability ranged from 44.4\% (V-MiniMax) to 88.9\% (V-Claude). Agentic-Kimi has 25 fixes instead of 27
because its second generation run for issues 25057 and 45912 is absent
from the source benchmark.

\begin{table}[h]
  \centering
  \footnotesize
  \setlength{\tabcolsep}{3pt}
  \caption{Resolution (Res.) and applicability (App.) rate (\%) per
  configuration and executor. Each configuration has 27 fixes
  (Agentic-Kimi: 25). Backbones: Gem.\ = Gemini 3.1 Pro, Cla.\ = Claude
  Opus 4.6, MiniM.\ = MiniMax M2.7. Executors: S5 = Sonnet~5,
  OC = OpenCUA-72B, MG = Muse Glimmer.}
  \label{tab:rq1-configs}
  \begin{tabular}{@{}l l rrrrrr rr rr rr@{}}
    \toprule
    & & \multicolumn{6}{c}{\textbf{Vanilla}} & \multicolumn{2}{c}{\textbf{Ablation}}
      & \multicolumn{2}{c}{\textbf{RAG}} & \multicolumn{2}{c}{\textbf{Agentic}} \\
    \cmidrule(lr){3-8}\cmidrule(lr){9-10}\cmidrule(lr){11-12}\cmidrule(l){13-14}
    & & \textbf{Gem.} & \textbf{GPT} & \textbf{Cla.} & \textbf{GLM} & \textbf{MiniM.} & \textbf{Kimi}
      & \textbf{Gem.} & \textbf{Kimi} & \textbf{Gem.} & \textbf{Kimi} & \textbf{Gem.} & \textbf{Kimi} \\
    \midrule
    \multirow{3}{*}{\textbf{Res.}}
      & S5 & 44.4 & 40.7 & \textbf{74.1} & 44.4 & 25.9 & 44.4 & 59.3 & 51.9 & 48.1 & 44.4 & 63.0 & 56.0 \\
      & OC & 51.9 & 48.1 & 51.9 & 37.0 & 44.4 & 33.3 & 44.4 & 40.7 & 48.1 & \textbf{55.6} & \textbf{55.6} & 36.0 \\
      & MG & 11.1 & 11.1 & 25.9 & 18.5 & 11.1 &  0.0 & 14.8 & 11.1 &  7.4 & 11.1 & 25.9 & \textbf{28.0} \\
    \addlinespace
    \multirow{3}{*}{\textbf{App.}}
      & S5 &  55.6 &  55.6 &  \textbf{88.9} &  51.9 &  44.4 &  66.7 &  70.4 &  74.1 &  63.0 &  70.4 &  81.5 &  72.0 \\
      & OC & \textbf{100.0} &  96.3 & \textbf{100.0} & \textbf{100.0} & \textbf{100.0} & \textbf{100.0} & \textbf{100.0} &  96.3 &  96.3 & \textbf{100.0} & \textbf{100.0} & \textbf{100.0} \\
      & MG &  63.0 &  63.0 &  \textbf{81.5} &  70.4 &  63.0 &  \textbf{81.5} &  66.7 &  77.8 &  70.4 &  70.4 &  74.1 &  64.0 \\
    \bottomrule
  \end{tabular}
\end{table}

\paragraph{Pipeline variants.}
Gemini~3.1 Pro and Kimi~K2.5 are the only backbones instantiated in all
four pipelines (Section~\ref{sec:llm-configs}), so they allow comparing
pipelines while holding the backbone fixed. With Gemini, Agentic had the highest resolution rate under both Sonnet~5 (63.0\%) and OpenCUA-72B (55.6\%); the lowest was Vanilla under Sonnet~5 (44.4\%) and Ablation under OpenCUA-72B (44.4\%). With Kimi, Agentic had the highest rate under Sonnet~5 (56.0\%) and RAG under OpenCUA-72B (55.6\%); no pipeline resolved fewer fixes than Vanilla (44.4\% under Sonnet~5, 33.3\% under OpenCUA-72B). Agentic also had the highest rate under Muse Glimmer for both backbones (25.9\% and 28.0\%).

\paragraph{Per subclass.}
Table~\ref{tab:rq1-subclass} breaks the results down by subclass. The subclass with the highest resolution rate differed by executor: External System \& Dependency under Sonnet~5 (59.3\%) and Muse Glimmer (21.3\%), and Faulty Configuration under OpenCUA-72B (65.7\%). Faulty Configuration fixes reached only 13.5\% under Muse Glimmer. The single Wrong Version issue had the lowest resolution rate under OpenCUA-72B (8.3\%) and Muse Glimmer (0.0\%), but not under Sonnet~5 (50.0\%), which completes the update and the gradient selection that this issue requires. Applicability was lowest for the Wrong Version issue under Sonnet~5 (44.4\%) and for External System under Muse Glimmer (67.6\%), and close to 100\% for OpenCUA-72B in every subclass.

\begin{table}[h]
  \centering
  \footnotesize
  \caption{Resolution and applicability rate (\%) per subclass.}
  \label{tab:rq1-subclass}
  \begin{tabular}{@{}l r r r r r r r r@{}}
    \toprule
    & & & \multicolumn{2}{c}{\textbf{Sonnet 5}} & \multicolumn{2}{c}{\textbf{OpenCUA-72B}} & \multicolumn{2}{c}{\textbf{Muse Glimmer}} \\
    \cmidrule(lr){4-5}\cmidrule(lr){6-7}\cmidrule(l){8-9}
    \textbf{Subclass} & \textbf{Iss.} & \textbf{Fixes} & \textbf{Res.} & \textbf{App.} & \textbf{Res.} & \textbf{App.} & \textbf{Res.} & \textbf{App.} \\
    \midrule
    Faulty Configuration & 5 & \textbf{178} & 43.8 &  66.3 & \textbf{65.7} &  98.9 & 13.5 & 69.7 \\
    Wrong Version        & 1 &  36 & 50.0 &  44.4 &  8.3 & \textbf{100.0} &  0.0 & \textbf{83.3} \\
    External System      & 3 & 108 & \textbf{59.3} &  \textbf{73.1} & 25.0 &  99.1 & \textbf{21.3} & 67.6 \\
    \bottomrule
  \end{tabular}
\end{table}

\begin{tcolorbox}[colback=gray!5,colframe=gray!60,boxrule=0.5pt,left=4pt,right=4pt,top=2pt,bottom=2pt]
\textbf{Answer to RQ1.} Depending on the executor, 14.6\% to 49.7\% of the 322
LLM-generated no-code fixes resolved the reported bug, and 66.1\% to
99.1\% were applicable. V-Claude had the highest resolution rate under
Sonnet~5 (74.1\%), but ranked third under OpenCUA-72B, where RAG-Kimi and Agentic-Gemini tied for first, and second under Muse Glimmer, where Agentic-Kimi led. The weakest
configuration and the best-resolved subclass also depended on the
executor. The Wrong Version issue was resolved least often under
OpenCUA-72B and Muse Glimmer, but not under Sonnet~5.
\end{tcolorbox}

\subsection{RQ2: Sensitivity to the Choice of Executor}
\label{sec:results-rq2}

\paragraph{Run-to-run stability.}
Table~\ref{tab:rq2-runs} reports the pass counts of each individual executor
run and the number of fixes resolved in at least one run. Within an executor, the pass rate varied by at most 3.4\% across runs (Sonnet~5, 38.8\% to 42.2\%), while Muse Glimmer trailed the other two executors by at least 20.5 points in every run. All three runs reached the same verdict for 92.2\% of fixes under Muse Glimmer, 82.3\% under Sonnet~5, and 73.9\% under OpenCUA-72B. Counting a fix as resolved if any run passed added 24 fixes over the best single run for Sonnet~5 (160 against 136), 38 for OpenCUA-72B (147 against 109), and 12 for Muse Glimmer (47 against 35). Under a majority vote the rates were 39.8\%, 30.7\%, and 9.9\%, so Sonnet~5 leads OpenCUA-72B under both aggregation rules.

\begin{table}[h]
  \centering
  \footnotesize
  \caption{Resolved fixes per individual executor run, fixes resolved in at
  least one run, and the share of fixes on which all three runs agreed
  ($n = 322$).}
  \label{tab:rq2-runs}
  \setlength{\tabcolsep}{4pt}
  \begin{tabular}{@{}l r r r r r@{}}
    \toprule
    \textbf{Executor} & \textbf{Run 1} & \textbf{Run 2} & \textbf{Run 3} & \textbf{Any run} & \textbf{All runs agree} \\
    \midrule
    Claude Sonnet 5 & \textbf{136 (42.2\%)} & 125 (38.8\%) & 130 (40.4\%) & \textbf{160 (49.7\%)} & 82.3\% \\
    OpenCUA-72B     & 100 (31.1\%) & \textbf{109 (33.9\%)} & 100 (31.1\%) & 147 (45.7\%) & 73.9\% \\
    Muse Glimmer    &  34 (10.6\%) &  \textbf{35 (10.9\%)} &  32 (9.9\%)  &  47 (14.6\%) & \textbf{92.2\%} \\
    \bottomrule
  \end{tabular}
\end{table}

\paragraph{Agreement between executors.}
All three executors reached the same verdict on 46.9\% of fixes. Sonnet~5 and OpenCUA-72B agreed on 62.4\% of fixes (Cohen's $\kappa = 0.25$), Sonnet~5 and Muse Glimmer on 64.3\% ($\kappa = 0.28$), and OpenCUA-72B and Muse Glimmer on 67.1\% ($\kappa = 0.30$). The ranking of the 12 configurations was only weakly preserved: Kendall's $\tau_b$ was 0.13 between Sonnet~5 and OpenCUA-72B, 0.45 between Sonnet~5 and Muse Glimmer, and 0.05 between OpenCUA-72B and Muse Glimmer. For a single configuration, the resolution rate differed across executors by 38.8\% on average (range 25.9 to 48.1; 95\% bootstrap interval 31.2 to 52.1). The disagreement is not concentrated on the three version-dependent issues, where the outcome depends on whether the executor performs the update: without them, agreement drops further ($\kappa$ = 0.22, 0.24, and 0.17), and all three executors agree on only 40.7\% of the fixes.

\paragraph{Per-issue differences.}
Table~\ref{tab:rq2-issues} lists resolved fixes per issue. On the three version-dependent issues, Sonnet~5 resolved the most fixes (16, 18, and 29 for 22252, 25379, and 31108), and the executors no longer resolved the same number of fixes. On 25379, OpenCUA-72B resolved 3 fixes and Muse Glimmer none, because an update alone is not enough there and a gradient must be chosen afterwards. On issue 24747, Sonnet~5 resolved 19 fixes, while neither OpenCUA-72B nor Muse Glimmer resolved any. On issue 37372, OpenCUA-72B resolved all 36 fixes against 20 for Sonnet~5 and 12 for Muse Glimmer, and on 20537 and 25057 it resolved roughly twice as many fixes as Sonnet~5. Muse Glimmer resolved no fix for issues 24747, 25057, 25379, and 45912.

\begin{table}[h]
  \centering
  \footnotesize
  \setlength{\tabcolsep}{4pt}
  \caption{Resolved fixes per issue and executor. Column headers are Brave
  GitHub issue numbers, grouped by subclass.}
  \label{tab:rq2-issues}
  \begin{tabular}{@{}l rrrrr r rrr r@{}}
    \toprule
    & \multicolumn{5}{c}{\textbf{Faulty Configuration}}
    & \textbf{\makecell{Wrong\\Version}}
    & \multicolumn{3}{c}{\textbf{External System}} & \\
    \cmidrule(lr){2-6}\cmidrule(lr){7-7}\cmidrule(lr){8-10}
    \textbf{Issue}
    & \textbf{20537} & \textbf{22987} & \textbf{25057} & \textbf{37372} & \textbf{45912}
    & \textbf{25379}
    & \textbf{22252} & \textbf{24747} & \textbf{31108} & \textbf{Total} \\
    \midrule
    Fixes        & 36 & 36 & 35 & 36 & 35 & 36 & 36 & 36 & 36 & 322 \\
    \midrule
    Sonnet 5     &  9 & 13 & 10 & 20 & \textbf{26} & \textbf{18} & \textbf{16} & \textbf{19} & \textbf{29} & \textbf{160} \\
    OpenCUA-72B  & \textbf{19} & \textbf{20} & \textbf{19} & \textbf{36} & 23 &  3 & 12 &  0 & 15 & 147 \\
    Muse Glimmer &  3 &  9 &  0 & 12 &  0 &  0 & 10 &  0 & 13 &  47 \\
    \bottomrule
  \end{tabular}
\end{table}

\paragraph{Episode termination.}
Each episode ends with the executor's own verdict, which is independent of
the checker. Sonnet~5 ended 83.0\% of its 966 episodes with an explicit
verdict (50.2\% DONE, 32.8\% BLOCKED), OpenCUA-72B 59.7\% (55.1\% DONE,
4.7\% BLOCKED), and Muse Glimmer 44.9\% (13.4\% DONE, 31.6\% BLOCKED);
nearly all remaining episodes exhausted the 20-step budget (16.3\%,
40.2\%, and 55.1\%). Across all executors, the checker recorded the bug as
resolved in 621 of the 1{,}146 episodes ending in DONE (54.2\%), 162 of the
1{,}077 episodes ending at the step budget (15.0\%), and 18 of the 667
episodes ending in BLOCKED (2.7\%).

\begin{tcolorbox}[colback=gray!5,colframe=gray!60,boxrule=0.5pt,left=4pt,right=4pt,top=2pt,bottom=2pt]
\textbf{Answer to RQ2.} The choice of executor changed the measured fix
capability substantially: overall resolution ranged from 14.6\% to
49.7\%, the rate of a single configuration differed across executors by
38.8\% on average, and the executors reached the same
verdict on only 46.9\% of the fixes. Each executor was stable across its
own runs (at most 3.4 points), so the differences between executors
exceeded their run-to-run variation. Configuration rankings were only
weakly preserved ($\tau_b$ between 0.05 and 0.45), and V-Claude, first
under Sonnet~5, ranked third under OpenCUA-72B and second under Muse
Glimmer.
\end{tcolorbox}

\subsection{RQ3: Agreement with Human Execution}
\label{sec:results-rq3}
The two authors independently executed all 75 randomly sampled
fixes and reached the same outcome on 78.7\% of them (59 fixes, Cohen's
$\kappa = 0.54$). Table~\ref{tab:rq3} reports, for each executor, the
percentage of sampled fixes on which its verdict matched the outcome of
each human executor and of the consensus set, with 95\% Wilson
confidence intervals. Claude Sonnet~5 matched the consensus outcome on 88.1\% of fixes ($\kappa = 0.74$), Muse Glimmer on 79.7\% ($\kappa = 0.48$), and OpenCUA-72B on 66.1\% ($\kappa = 0.28$).
 
The mismatches differed in direction. Of the 20 consensus mismatches of OpenCUA-72B, 13 were fixes it counted as resolved that the humans did not. Of the 12 mismatches of Muse Glimmer, 10 were fixes it did not resolve although the humans did, and Sonnet~5 erred in both directions (5 false positives and 2 false negatives). On the 75 sampled fixes, OpenCUA-72B resolved 33, Muse Glimmer 13, and Sonnet~5 33, against 25 and 29 for the two humans. Replacing ``any run'' with a majority vote over the three runs changed the agreement with the consensus to 86.4\% for Sonnet~5, 72.9\% for OpenCUA-72B, and 81.4\% for Muse Glimmer, so the aggregation rule did not alter the ordering. The human outcomes differed on the remaining 16 fixes, which are therefore not part of the consensus set.
 
\begin{table}[h]
  \centering
  \footnotesize
  \caption{Agreement (\%) between each executor and human execution on the 75
  randomly sampled fixes, with 95\% Wilson confidence intervals.
  Consensus: the 59 fixes on which both humans reached the same outcome.}
  \label{tab:rq3}
  \setlength{\tabcolsep}{4pt}
  \begin{tabular}{@{}l c c c c@{}}
    \toprule
    \textbf{Executor} & \textbf{Human 1} & \textbf{Human 2}
                 & \textbf{Consensus} & \textbf{$\kappa$ (consensus)} \\
    \midrule
    Claude Sonnet 5 & 81.3 {\scriptsize[71.1, 88.5]} & 78.7 {\scriptsize[68.1, 86.4]} & \textbf{88.1 {\scriptsize[77.5, 94.1]}} & \textbf{0.74} \\
    OpenCUA-72B     & 68.0 {\scriptsize[56.8, 77.5]} & 57.3 {\scriptsize[46.1, 67.9]} & 66.1 {\scriptsize[53.4, 76.9]} & 0.28 \\
    Muse Glimmer    & 78.7 {\scriptsize[68.1, 86.4]} & 68.0 {\scriptsize[56.8, 77.5]} & 79.7 {\scriptsize[67.7, 88.0]} & 0.48 \\
    \midrule
    Human 1 vs.\ Human 2 & \multicolumn{2}{c}{78.7 {\scriptsize[68.1, 86.4]}} & & 0.54 \\
    \bottomrule
  \end{tabular}
\end{table}
 
\begin{tcolorbox}[colback=gray!5,colframe=gray!60,boxrule=0.5pt,left=4pt,right=4pt,top=2pt,bottom=2pt]
\textbf{Answer to RQ3.} The authors agreed on 78.7\% of the
75 sampled fixes ($\kappa = 0.54$). On the fixes with a shared human
outcome, Sonnet~5 matched on 88.1\% ($\kappa = 0.74$), Muse Glimmer on
79.7\% ($\kappa = 0.48$), and OpenCUA-72B on 66.1\% ($\kappa = 0.28$).
OpenCUA-72B and Sonnet~5 overstated and Muse Glimmer understated the
number of resolved fixes; Sonnet~5 had both the highest resolution rate
and the highest agreement with the humans.
\end{tcolorbox}

\section{Discussion}
\label{sec:discussion}

\subsection{The Executor Is Part of the Measurement}
\label{sec:discussion-executor}

RQ2 shows that an executor-based evaluation does not measure fix quality in
isolation. The same 322 fixes produced resolution rates between 14.6\% and 49.7\%, and the rate of a single configuration moved by 38.8\% on average when only the executor changed. These gaps are much larger than the run-to-run variation within any executor (at most 3.4 points), so they reflect systematic properties of the executors rather
than noise. Conclusions about which LLM produces better no-code fixes are
therefore conditional on the executor. Even the strongest configuration was not stable: V-Claude led under Sonnet~5 but ranked third under OpenCUA-72B and second under Muse Glimmer, and the rankings of the 12 configurations were nearly unrelated between Sonnet~5 and OpenCUA-72B ($\tau_b = 0.13$) and between OpenCUA-72B and Muse Glimmer ($\tau_b = 0.05$). Some comparisons changed substantially: RAG-Kimi resolved 22.3\% more fixes than V-Kimi under OpenCUA-72B (55.6\% against 33.3\%) and exactly as many under Sonnet~5.

The executors also differed in where they succeeded. OpenCUA-72B resolved
every fix for issue 37372\footnote{\url{https://github.com/brave/brave-browser/issues/37372}}, whose checker is lenient (Section~\ref{sec:threats}), while Sonnet~5 resolved 20 of 36, whereas
Sonnet~5 was the only executor to resolve any fix for issue 24747\footnote{\url{https://github.com/brave/brave-browser/issues/24747}}. Muse
Glimmer exhausted its step budget in 55.1\% of its episodes and resolved
no fix for four of the nine issues. No single executor is thus a neutral
instrument, and a verdict of ``fix failed'' can mean that the executor
failed. We draw three practical implications for execution-based
evaluation with agentic executors. First, the executor and its harness
should be reported as part of the experimental setup, in the same way as
the model under evaluation. Second, results should be obtained with more
than one executor or calibrated against human execution, as we do in RQ3.
Third, run-level results and the aggregation rule should be reported,
because the rule interacts with executor stability: counting a fix as
resolved if any run passed added 38 fixes over the best single run for
OpenCUA-72B, the least stable executor, but only 12 to 24 for the other two.

RQ3 indicates which of the three executors agrees best with human execution. Sonnet~5, which also had the highest resolution rate (49.7\%), agreed most with the humans (88.1\% with the consensus, $\kappa = 0.74$), although it resolved more fixes than the humans on the sample (33 against 25 and 29). OpenCUA-72B had the lowest agreement (66.1\%, $\kappa = 0.28$), and most of its mismatches were false positives (13 of 20). Muse Glimmer mismatched in the opposite direction (10 of its 12 mismatches were fixes it did not resolve although the humans did), which is consistent with the 55.1\% of its episodes that exhausted the step budget. Rates under OpenCUA-72B and, to a lesser extent, Sonnet~5 are therefore likely to overstate, and those under Muse Glimmer to understate, the share of fixes that work when a human applies them.

The executors' termination status also shows why the verdict should not be
left to the executor. An executor reports DONE when it believes it has applied
the fix, not when the bug is gone, and the two diverge in both
directions: the checker found the bug resolved in only 621 of the 1{,}146 episodes that ended in DONE, while 162 of the 1{,}077 episodes that ran out of steps nevertheless left the browser in a resolved state. On the version-dependent issues alone, Sonnet~5, OpenCUA-72B, and Muse Glimmer said DONE while the checker failed in 7, 121, and 1 of 324 episodes, and the checker passed without a DONE report in 54, 38, and 26. Using
the agent's own completion signal as the outcome would therefore both
inflate and deflate the measured capability, which is consistent with
the observation of Hong et al.~\cite{hong2026diageval} that GUI agents
often misattribute their own execution failures. Separating the executor
from an independent checker keeps the verdict reproducible, at the cost
of per-issue setup and checker code (Section~\ref{sec:threats}).

\subsection{What Text Similarity Misses}
\label{sec:discussion-bertscore}

As argued in Section~\ref{sec:intro} reference-based metrics cannot show
whether a no-code fix works. Our execution results allow a first look at
this claim. For each LLM-generated fix, we took the
BERTScore~\cite{zhang2019bertscore} value against the maintainer's
ground-truth fix released in the replication package of G\"on et
al.~\cite{gon2026automatedrootcausesubclassificationnocode}, and examined
the 108 fixes that failed in all nine attempts, that is, in every run
under every executor. Because no executor resolved these fixes, their
failure does not depend on the executor sensitivity discussed in
Section~\ref{sec:discussion-executor}.

These fixes received BERTScore F1 values between 0.757 and 0.867 (median 0.831), within or above the range of 0.81 to 0.82 that G\"on et
al.~\cite{gon2026automatedrootcausesubclassificationnocode} report for
fixes across their benchmark. A fix that never removed the reported
symptom was thus scored as close to the maintainer's resolution as a
typical fix. The 214 fixes that at least one executor resolved scored the same (median 0.832, mean 0.824 against 0.823). Scores also varied far more between issues than within them.
Fixes for issue 25379 averaged 0.762, while those for every other issue with such fixes averaged between 0.818 and 0.846 (issue 37372 has none, because OpenCUA-72B resolved all its fixes), which suggests that the score reflects
the wording of the reference resolution more than the correctness of the
fix. A single BERTScore threshold therefore cannot separate working from
failing fixes across issues, and a comparison of configurations by
BERTScore partly reflects which issues their fixes belong to.

Execution-based evaluation does not replace text-based evaluation
everywhere. It applies only to reports whose bug can be re-established in
a controlled environment (Section~\ref{sec:discussion-scarcity}), whereas
BERTScore and an LLM judge apply to any report with a reference fix. For
reports that can be executed, however, these fixes show that a high
similarity to the maintainer's resolution does not indicate that a fix
works.

\subsection{What the Results Say About No-Code Fix Generation}
\label{sec:discussion-fixes}

Even under the executor with the highest resolution rate and the highest agreement with the humans (Sonnet~5, 49.7\%), just under half of the LLM-generated fixes resolved the reported bug (39.8\% under a majority vote), and the best configuration reached 74.1\%. Automatically generated no-code fixes are therefore not
yet reliable enough to be sent to reporters without review, although the
best configurations resolve a majority of cases under two of the three
executors. No configuration led under every executor. V-Claude, which also had the highest applicability under Sonnet~5 (88.9\%), and Agentic-Gemini were the only configurations in the top three under all three executors, but the intervals overlap heavily, so we do not claim a robust winner. The pipeline enhancements proposed by Gön et al.~\cite{gon2026automatedrootcausesubclassificationnocode} did not translate into consistent gains: for Gemini, RAG resolved fewer fixes than Vanilla under OpenCUA-72B and Muse Glimmer but more under Sonnet~5, whereas for Kimi neither RAG nor Agentic web search resolved fewer fixes than Vanilla under Sonnet~5 or OpenCUA-72B. Agentic web search had the highest rate for both backbones under Sonnet~5 and Muse Glimmer, but not for Kimi under OpenCUA-72B.

Applicability and resolution capture different ways in which a fix can
fail. A configuration can fail by not proposing an executable action, as
V-MiniMax did for almost half of its fixes under Sonnet~5, or by
proposing an action that does not remove the symptom. The two failure
modes call for different remedies: the first concerns the model's
willingness to commit to a concrete instruction, the second its knowledge
of the application. At the same time, applicability as measured here
depends on the executor, since OpenCUA-72B rarely declared any fix
non-applicable.

\subsection{The Scarcity of Executable No-Code Fixes}
\label{sec:discussion-scarcity}

The reduction from 302 annotated reports to 51 candidates and finally to
nine executable issues is worth reading as a finding rather than merely as
a description of our sample. Bug reports resolved without a code change
are only part of what an issue tracker receives~\cite{aktas2023issue,erfani2014works},
and only a subset of those are resolved by an action that can be carried out in an
environment at all; the remainder are answered with an explanation, a
request for further information, or a decision that the reported behavior
is intended~\cite{erfani2014works} (Section~\ref{sec:dataset}). What makes such reports costly to collect is that this
distinction leaves no structural trace. Issue trackers do not label
invalid reports as such, and their resolutions are not linked to commits,
so there is no signal analogous to a fixing commit that can be mined
automatically, as is routinely done when constructing code-level
benchmarks. Identifying a no-code fix requires reading the maintainer
discussion of a closed issue and judging what the resolution consisted
of. Any effort to scale execution-based evaluation of no-code fixes will
therefore have to confront dataset construction as a research problem in
its own right, most plausibly by automating the recognition of no-code
resolutions in maintainer comments.

The reduction observed before and at the sanity gates is informative for
a second reason. Of the 51 candidates, 38 could not be set up in a Linux
browser environment at all, because they depended on another platform, a
live backend, or information the report did not provide~\cite{bruno2026writing}, and 4 more did
not exhibit the reported bug when tested. These are reports whose resolution was
accepted by both the reporter and the maintainer, yet whose bug-fix pair
could not be re-established outside the original reporting context. This
suggests that no-code bugs are frequently entangled with state that the
report does not capture, such as an accumulated user profile, a
particular operating system or hardware configuration, the behavior of a
third-party service at the time of reporting, or a server-side change
that no local rollback can undo. Seen this way, the gates are not only a
filter that reduced our evaluation set but a measurement instrument: they
quantify how often a documented no-code fix can be independently
re-executed at all, which is a precondition that any execution-based
evaluation of this task, ours or otherwise, must satisfy. We consequently
kept them strict, on the view that a verdict grounded in a verified
buggy state and a verified fixed state carries more evidential weight
than a larger set of verdicts resting on unchecked assumptions about the
environment. Threats arising from the resulting set size are discussed in
Section~\ref{sec:threats}.

\section{Threats to Validity}
\label{sec:threats}

\subsection{Internal Validity}

\paragraph{Executor reliability.} Every verdict in our study depends on an 
executor carrying out a natural-language fix in a real browser. An executor may
misapply a fix, apply it only partially, or change unrelated state, in
which case a failed fix says nothing about the fix itself. The
20-step budget adds a related threat: an episode that is cut off may have
succeeded with more steps, which matters most for Muse Glimmer, whose
episodes exhausted the budget in 55.1\% of cases. To reduce the effect of
individual execution failures, we executed every fix three times per executor
and counted a fix as resolved if any run succeeded, and we compare executor
outcomes with human execution on a random sample of fixes. Executors could also act outside the browser: the container desktop offers a shell, and 10 OpenCUA-72B episodes typed shell-like commands. Nine of them failed without an update and one passed through the settings-page update, so no resolved episode depended on a shell.

\paragraph{Harness differences.} The three executors could not be driven by an
identical harness. Sonnet~5 used its native computer-use tool,
OpenCUA-72B used its own action format with the screenshot history window
of its reference harness, and Muse Glimmer, which has no official
computer-use harness, was driven through a function-calling schema that
we designed. The executors also used different sampling settings
(Section~\ref{sec:cuas}), which we did not tune. A higher temperature
yields more varied runs and thus benefits more from counting a fix as
resolved if any of three runs succeeds.
Differences between executors may therefore partly stem from
their harnesses and sampling settings, and our own schema may disadvantage Muse Glimmer. We
kept the task prompt, the step budget, and the environment identical
across executors, and we interpret RQ2 as a comparison of executors together with
their harnesses rather than of the underlying models alone.

\paragraph{Checker validity.} Whether a fix resolves the bug is decided
by an issue-specific checker that reads the browser state tied to the
symptom in the bug report, such as a preference value or a property of
the rendered page. A checker may accept a state that was reached without
addressing the reported problem, or reject a valid fix that removes the
symptom through a different route. In particular, for issues 25057 and
37372, the expected value is the browser's default, which the browser does
not store explicitly, so the checker also accepts the absence of the
preference; any action that removes it, including a general reset of
the browser settings, therefore counts as a resolution. We derived each
checker from the actual result stated in the bug report, release all
checkers and execution traces in our replication package, and use the
human execution in RQ3 as an independent check of their verdicts.

\paragraph{Metric definitions.} Two of our metrics involve design
choices that affect the reported values. Applicability relies on the
executor's decision to declare a fix BLOCKED and thus reflects the executor's
judgment as well as the fix; OpenCUA-72B, for instance, ended only 4.7\%
of its episodes in BLOCKED. Counting a fix as resolved if any of three
runs succeeded yields higher rates than a majority vote, and the gain is
larger for less stable executors. We therefore report every individual
run (Table~\ref{tab:rq2-runs}) and provide majority-vote results in the
replication package. On the RQ3 sample, a majority vote changed the agreement with the human consensus by at most 6.8\% per executor, and dropping the six fixes on which the first human recorded partial or not applicable changed the agreement by at most 4.2 points. The BERTScore analysis in Section~\ref{sec:discussion-bertscore} covers only
fixes that failed in every attempt, whereas the range reported by G\"on et
al.~\cite{gon2026automatedrootcausesubclassificationnocode} spans all seven
subclasses and their reference fixes.

\paragraph{Buggy-state fidelity.} Per-fix episodes start from a
reconstructed buggy state: a fresh profile on the gate-verified build,
with the buggy value pre-seeded for three issues, or the reported build
for version-type issues. The sanity gates confirm that this state shows
the reported symptom and that the maintainer's fix removes it, but the
reporter's original state may have depended on context we cannot
recreate, such as an accumulated profile or a server-side condition. Two
reports state no version (25057 and 37372), so we used a nearby pinned
build for them.

\paragraph{Human reference.} The human executors in RQ3 are the first two
authors. Their familiarity with the pipeline, and in some cases with a
bug's ground-truth resolution, may influence how they execute a fix.
We blinded both executors to the configuration that generated each fix and
to the executors' outcomes, and had both execute every sampled fix
independently. Their agreement was moderate (78.7\%, $\kappa = 0.54$), and
the second human resolved more fixes than the first (29 against 25), so the
human outcome is a reference rather than a validated ground truth.
Restricting the comparison to the 59 fixes with a shared outcome removes
the fixes that were hardest to judge and may therefore overstate executor
accuracy; we report the agreement with each human separately for this
reason. Each human executed a fix once, whereas an executor's verdict
aggregates three runs, and the consensus set contains only 19 resolved
fixes, so the intervals in Table~\ref{tab:rq3} are wide and we do not test
differences between executors for significance.

\subsection{External Validity}

\paragraph{Application and bug types.} All issues come from a single
application, the Brave browser, and a single prior
benchmark~\cite{gon2026automatedrootcausesubclassificationnocode}, and
are restricted to the three subclasses whose resolutions are executable
actions. Our findings may not transfer to other browsers, to desktop or
mobile applications with different interaction models, or to the
subclasses we excluded by design. Moreover, our environment runs on
Linux, and 17 of the 51 candidate issues were excluded because they were
specific to another platform; the evaluated issues may therefore differ
systematically from those that could not be reproduced.

\paragraph{Size of the evaluation set.} The effective evaluation set
contains nine issues and 322 fixes, further divided across 12
configurations and three executors. The subclass breakdown rests on five, one,
and three issues, so the Wrong Version result describes a single issue.
We report agreement statistics and rank correlations descriptively, together with issue-clustered bootstrap intervals, and do not perform significance tests. The intervals are wide and overlap (for example 38.6 to 61.3 for Sonnet~5 and 27.6 to 64.5 for OpenCUA-72B), so differences between configurations and between executors should be read as descriptive rather than statistically confirmed.

\section{Conclusion}
\label{sec:conclusion}

We presented a rollback-based, automated execution-driven pipeline that verifies
LLM-generated no-code fixes by applying them with an executor in
a real browser and checking the resulting state. Applied to 322 fixes
from 12 LLM-based configurations on nine reproducible Brave bug reports,
the pipeline shows that current LLMs resolve between 14.6\% and 49.7\% of the reported bugs depending on the executor, with V-Claude the strongest configuration under Sonnet~5 but not under the other two executors. The choice of executor shifted the measured capability of a configuration by 38.8\% on average and only weakly preserved configuration rankings, which makes the
executor a first-order variable in any execution-based evaluation of
this kind. Only nine of 51 candidate reports could be
re-established in a controlled environment, so scaling this form of
evaluation will require automated construction of executable no-code
benchmarks. Compared with human execution on a random sample, Sonnet~5 was the most accurate executor (88.1\% agreement with the human consensus), whereas OpenCUA-72B overstated resolution and agreed on 66.1\%. Future work will extend the pipeline to more applications
and bug types, and study how executor harnesses can be standardized so
that verdicts become comparable across executors.

\section{Data Availability}
\label{sec:data}
Our replication package is available at
\url{https://figshare.com/s/eda4e478ccbf242fbdab}. It contains the
Dockerfile and build scripts for the pinned Brave builds of the evaluated issues, the per-issue environment setup (profile seeds and
scripted gate actions), the issue-specific checkers, the executor prompts
and action interfaces (including our function-calling schema for Muse
Glimmer), the verdicts of executor episodes and the text
traces of them, the 322 evaluated fix texts, the human verdicts and sampling seed
for RQ3, the status of all 51 candidate issues with the reason each
excluded issue was dropped, and the scripts that produce every table,
including the majority-vote variants.

\section*{Acknowledgements}
We gratefully acknowledge OpenZeka for providing access to the GPU resources that made the experiments in this study possible. This work was funded by TÜBİTAK (The Scientific and Technological Research Council of Turkey) through the 1001 Scientific and Technological Research Projects Funding Program under Project No. 125E371.

\newpage
\bibliographystyle{ACM-Reference-Format}
\bibliography{references}

@misc{gon2026automatedrootcausesubclassificationnocode,
      title={Automated Root-Cause Subclassification and No-Code Fix Generation for Invalid Bug Reports}, 
      author={Mahmut Furkan Gon and Emre Dinc and Tevfik Emre Sungur and Eray Tuzun},
      year={2026},
      eprint={2605.17561},
      archivePrefix={arXiv},
      primaryClass={cs.SE},
      url={https://arxiv.org/abs/2605.17561}, 
}

@misc{torun2026evaluationllmbasedsoftwareengineering,
      title={Evaluation of LLM-Based Software Engineering Tools: Practices, Challenges, and Future Directions}, 
      author={Utku Boran Torun and Veli Karakaya and Ali Babar and Eray Tüzün},
      year={2026},
      eprint={2604.24621},
      archivePrefix={arXiv},
      primaryClass={cs.SE},
      url={https://arxiv.org/abs/2604.24621}, 
}

@article{jiang2026survey,
  title={A survey on large language models for code generation},
  author={Jiang, Juyong and Wang, Fan and Shen, Jiasi and Kim, Sungju and Kim, Sunghun},
  journal={ACM Transactions on Software Engineering and Methodology},
  volume={35},
  number={2},
  pages={1--72},
  year={2026},
  publisher={ACM New York, NY}
}

@article{zhang2023planning,
  title={Planning with large language models for code generation},
  author={Zhang, Shun and Chen, Zhenfang and Shen, Yikang and Ding, Mingyu and Tenenbaum, Joshua B and Gan, Chuang},
  journal={arXiv preprint arXiv:2303.05510},
  year={2023}
}

@article{jiang2024self,
  title={Self-planning code generation with large language models},
  author={Jiang, Xue and Dong, Yihong and Wang, Lecheng and Fang, Zheng and Shang, Qiwei and Li, Ge and Jin, Zhi and Jiao, Wenpin},
  journal={ACM Transactions on Software Engineering and Methodology},
  volume={33},
  number={7},
  pages={1--30},
  year={2024},
  publisher={ACM New York, NY}
}

@article{pornprasit2024fine,
  title={Fine-tuning and prompt engineering for large language models-based code review automation},
  author={Pornprasit, Chanathip and Tantithamthavorn, Chakkrit},
  journal={Information and Software Technology},
  volume={175},
  pages={107523},
  year={2024},
  publisher={Elsevier}
}

@inproceedings{cihan2025automated,
  title={Automated code review in practice},
  author={Cihan, Umut and Haratian, Vahid and {\.I}{\c{c}}{\"o}z, Arda and G{\"u}l, Mert Kaan and Devran, {\"O}mercan and Bayendur, Emircan Furkan and U{\c{c}}ar, Baykal Mehmet and T{\"u}z{\"u}n, Eray},
  booktitle={2025 IEEE/ACM 47th International Conference on Software Engineering: Software Engineering in Practice (ICSE-SEIP)},
  pages={425--436},
  year={2025},
  organization={IEEE}
}

@article{yu2024fine,
  title={Fine-tuning large language models to improve accuracy and comprehensibility of automated code review},
  author={Yu, Yongda and Rong, Guoping and Shen, Haifeng and Zhang, He and Shao, Dong and Wang, Min and Wei, Zhao and Xu, Yong and Wang, Juhong},
  journal={ACM transactions on software engineering and methodology},
  volume={34},
  number={1},
  pages={1--26},
  year={2024},
  publisher={ACM New York, NY}
}

@article{schafer2023empirical,
  title={An empirical evaluation of using large language models for automated unit test generation},
  author={Sch{\"a}fer, Max and Nadi, Sarah and Eghbali, Aryaz and Tip, Frank},
  journal={IEEE Transactions on Software Engineering},
  volume={50},
  number={1},
  pages={85--105},
  year={2023},
  publisher={IEEE}
}

@article{celik2025review,
  title={A review of large language models for automated test case generation},
  author={Celik, Arda and Mahmoud, Qusay H},
  journal={Machine Learning and Knowledge Extraction},
  volume={7},
  number={3},
  pages={97},
  year={2025},
  publisher={MDPI}
}

@article{zhang2024testbench,
  title={Testbench: Evaluating class-level test case generation capability of large language models},
  author={Zhang, Quanjun and Shang, Ye and Fang, Chunrong and Gu, Siqi and Zhou, Jianyi and Chen, Zhenyu},
  journal={arXiv preprint arXiv:2409.17561},
  year={2024}
}

@article{du2024llm,
  title={LLM-BRC: A large language model-based bug report classification framework},
  author={Du, Xiaoting and Liu, Zhihao and Li, Chenglong and Ma, Xiangyue and Li, Yingzhuo and Wang, Xinyu},
  journal={Software Quality Journal},
  volume={32},
  number={3},
  pages={985--1005},
  year={2024},
  publisher={Springer}
}

@inproceedings{aracena2024applying,
  title={Applying large language models to issue classification},
  author={Aracena, Gabriel and Luster, Kyle and Santos, Fabio and Steinmacher, Igor and Gerosa, Marco Aurelio},
  booktitle={Proceedings of the Third ACM/IEEE International Workshop on NL-based Software Engineering},
  pages={57--60},
  year={2024}
}

@article{colavito2025benchmarking,
  title={Benchmarking large language models for automated labeling: The case of issue report classification},
  author={Colavito, Giuseppe and Lanubile, Filippo and Novielli, Nicole},
  journal={Information and Software Technology},
  volume={184},
  pages={107758},
  year={2025},
  publisher={Elsevier}
}

@inproceedings{herzig2013its,
  title={It's not a bug, it's a feature: how misclassification impacts bug prediction},
  author={Herzig, Kim and Just, Sascha and Zeller, Andreas},
  booktitle={2013 35th international conference on software engineering (ICSE)},
  pages={392--401},
  year={2013},
  organization={IEEE}
}

@article{kang2024evaluating,
  title={Evaluating diverse large language models for automatic and general bug reproduction},
  author={Kang, Sungmin and Yoon, Juyeon and Askarbekkyzy, Nargiz and Yoo, Shin},
  journal={IEEE Transactions on Software Engineering},
  volume={50},
  number={10},
  pages={2677--2694},
  year={2024},
  publisher={IEEE}
}

@article{he2025code,
  title={From code to courtroom: Llms as the new software judges},
  author={He, Junda and Shi, Jieke and Zhuo, Terry Yue and Treude, Christoph and Sun, Jiamou and Xing, Zhenchang and Du, Xiaoning and Lo, David},
  journal={arXiv preprint arXiv:2503.02246},
  year={2025}
}

@inproceedings{yapaugci2025agents,
  title={Agents in the Sandbox: End-to-End Crash Bug Reproduction for Minecraft},
  author={Yapa{\u{g}}c{\i}, Eray and {\"O}zt{\"u}rk, Yavuz Alp Sencer and T{\"u}z{\"u}n, Eray},
  booktitle={2025 40th IEEE/ACM International Conference on Automated Software Engineering (ASE)},
  pages={3095--3107},
  year={2025},
  organization={IEEE}
}

@inproceedings{ir2026fixpad++,
  title={Fixpad++: Automated Bug Fix Verification Using LLM Agents},
  author={{\.I}r, Mustafa {\"O}zkan and Dedeler, Mehmet and Koyuncu, An{\i}l and T{\"u}z{\"u}n, Eray},
  booktitle={Proceedings of the 3rd ACM International Conference on AI-Powered Software},
  pages={248--258},
  year={2026}
}

@inproceedings{dincc2025judge,
  title={Judge the Votes: A System to Classify Bug Reports and Give Suggestions},
  author={Din{\c{c}}, Emre and T{\"u}z{\"u}n, Eray},
  booktitle={2025 2nd IEEE/ACM International Conference on AI-powered Software (AIware)},
  pages={01--10},
  year={2025},
  organization={IEEE}
}

@article{torun2026past,
  title={Past, present, and future of bug tracking in the generative ai era},
  author={Torun, Utku Boran and Demircan, Mehmet Taha and G{\"o}n, Mahmut Furkan and T{\"u}z{\"u}n, Eray},
  journal={ACM Transactions on Software Engineering and Methodology},
  year={2026},
  publisher={ACM New York, NY}
}

@inproceedings{jimenez2024swe,
  title={Swe-bench: Can language models resolve real-world github issues?},
  author={Jimenez, Carlos E and Yang, John and Wettig, Alexander and Yao, Shunyu and Pei, Kexin and Press, Ofir and Narasimhan, Karthik},
  booktitle={International Conference on Learning Representations},
  volume={2024},
  pages={54107--54157},
  year={2024}
}

@article{xie2024osworld,
  title={Osworld: Benchmarking multimodal agents for open-ended tasks in real computer environments},
  author={Xie, Tianbao and Zhang, Danyang and Chen, Jixuan and Li, Xiaochuan and Zhao, Siheng and Cao, Ruisheng and Hua, Toh J and Cheng, Zhoujun and Shin, Dongchan and Lei, Fangyu and others},
  journal={Advances in Neural Information Processing Systems},
  volume={37},
  pages={52040--52094},
  year={2024}
}

@inproceedings{zhou2024webarena,
  title={Webarena: A realistic web environment for building autonomous agents},
  author={Zhou, Shuyan and Xu, Frank F and Zhu, Hao and Zhou, Xuhui and Lo, Robert and Sridhar, Abishek and Cheng, Xianyi and Ou, Tianyue and Bisk, Yonatan and Fried, Daniel and others},
  booktitle={International Conference on Learning Representations},
  volume={2024},
  pages={15585--15606},
  year={2024}
}

@article{chen2025standing,
  title={Standing on the shoulders of giants: Bug-aware automated gui testing via retrieval augmentation},
  author={Chen, Mengzhuo and Liu, Zhe and Chen, Chunyang and Wang, Junjie and Wu, Boyu and Hu, Jun and Wang, Qing},
  journal={Proceedings of the ACM on Software Engineering},
  volume={2},
  number={FSE},
  pages={825--846},
  year={2025},
  publisher={ACM New York, NY, USA}
}

@inproceedings{antoniol2008bug,
  title={Is it a bug or an enhancement? a text-based approach to classify change requests},
  author={Antoniol, Giuliano and Ayari, Kamel and Di Penta, Massimiliano and Khomh, Foutse and Gu{\'e}h{\'e}neuc, Yann-Ga{\"e}l},
  booktitle={Proceedings of the 2008 conference of the center for advanced studies on collaborative research: meeting of minds},
  pages={304--318},
  year={2008}
}

@inproceedings{zanetti2013categorizing,
  title={Categorizing bugs with social networks: a case study on four open source software communities},
  author={Zanetti, Marcelo Serrano and Scholtes, Ingo and Tessone, Claudio Juan and Schweitzer, Frank},
  booktitle={2013 35th International Conference on Software Engineering (ICSE)},
  pages={1032--1041},
  year={2013},
  organization={IEEE}
}

@article{fan2018chaff,
  title={Chaff from the wheat: Characterizing and determining valid bug reports},
  author={Fan, Yuanrui and Xia, Xin and Lo, David and Hassan, Ahmed E},
  journal={IEEE transactions on software engineering},
  volume={46},
  number={5},
  pages={495--525},
  year={2018},
  publisher={IEEE}
}

@inproceedings{otoom2019automated,
  title={Automated classification of software bug reports},
  author={Otoom, Ahmed Fawzi and Al-jdaeh, Sara and Hammad, Maen},
  booktitle={Proceedings of the 9th international conference on information communication and management},
  pages={17--21},
  year={2019}
}

@article{pandey2017automated,
  title={Automated classification of software issue reports using machine learning techniques: an empirical study},
  author={Pandey, Nitish and Sanyal, Debarshi Kumar and Hudait, Abir and Sen, Amitava},
  journal={Innovations in Systems and Software Engineering},
  volume={13},
  number={4},
  pages={279--297},
  year={2017},
  publisher={Springer}
}

@inproceedings{meng2023bug,
  title={Which bug reports are valid and why? Using the BERT transformer to classify bug reports and explain their validity},
  author={Meng, Qianru and Visser, Joost},
  booktitle={Proceedings of the 4th European Symposium on Software Engineering},
  pages={52--60},
  year={2023}
}

@article{laiq2025comparative,
  title={A comparative analysis of ML techniques for bug report classification},
  author={Laiq, Muhammad and bin Ali, Nauman and B{\"o}rstler, J{\"u}rgen and Engstr{\"o}m, Emelie},
  journal={Journal of Systems and Software},
  volume={227},
  pages={112457},
  year={2025},
  publisher={Elsevier}
}

@inproceedings{feng2024prompting,
  title={Prompting is all you need: Automated android bug replay with large language models},
  author={Feng, Sidong and Chen, Chunyang},
  booktitle={Proceedings of the 46th IEEE/ACM International Conference on Software Engineering},
  pages={1--13},
  year={2024}
}

@article{kolthoff2025guispector,
  title={GUISpector: An MLLM Agent Framework for Automated Verification of Natural Language Requirements in GUI Prototypes},
  author={Kolthoff, Kristian and Kretzer, Felix and Ponzetto, Simone Paolo and Maedche, Alexander and Bartelt, Christian},
  journal={arXiv preprint arXiv:2510.04791},
  year={2025}
}

@article{hong2026diageval,
  title={DiagEval: Trajectory-Conditioned Diagnosis for Reliable Software Evaluation with GUI Agents},
  author={Hong, Sirui and Liu, Zhijie and Li, Tengfei and Tao, Wei and Wu, Yifan and Wu, Chenglin},
  journal={arXiv preprint arXiv:2605.17439},
  year={2026}
}

@book{cochran1977sampling,
  title={Sampling techniques},
  author={Cochran, William Gemmell},
  volume={3},
  number={4},
  year={1977},
  publisher={wiley New York}
}

@misc{meta2026museglimmer,
  author       = {{Meta Superintelligence Labs}},
  title        = {Introducing Muse Glimmer: An Open Agentic Model},
  year         = {2026},
  month        = aug,
  howpublished = {\url{https://research.meta.ai/blog/introducing-muse-glimmer-open-agentic-model}},
  note         = {Accessed: 2026-09-24}
}

@misc{anthropic2026sonnet5,
  author       = {{Anthropic}},
  title        = {Introducing {Claude Sonnet 5}},
  year         = {2026},
  month        = jun,
  howpublished = {\url{https://www.anthropic.com/news/claude-sonnet-5}},
  note         = {Accessed: 2026-09-24}
}

@misc{google2026gemini31pro,
  author       = {{Google}},
  title        = {{Gemini 3.1 Pro}: A Smarter Model for Your Most Complex Tasks},
  year         = {2026},
  month        = feb,
  howpublished = {\url{https://blog.google/innovation-and-ai/models-and-research/gemini-models/gemini-3-1-pro/}},
  note         = {Accessed: 2026-10-01}
}

@misc{openai2026gpt54,
  author       = {{OpenAI}},
  title        = {Introducing {GPT-5.4}},
  year         = {2026},
  month        = mar,
  howpublished = {\url{https://openai.com/index/introducing-gpt-5-4/}},
  note         = {Accessed: 2026-10-01}
}

@misc{anthropic2026opus46,
  author       = {{Anthropic}},
  title        = {Introducing {Claude Opus 4.6}},
  year         = {2026},
  month        = feb,
  howpublished = {\url{https://www.anthropic.com/news/claude-opus-4-6}},
  note         = {Accessed: 2026-10-01}
}

@article{zeng2026glm,
  title={Glm-5: from vibe coding to agentic engineering},
  author={Zeng, Aohan and Lv, Xin and Hou, Zhenyu and Du, Zhengxiao and Zheng, Qinkai and Chen, Bin and Yin, Da and Ge, Chendi and Huang, Chenghua and Xie, Chengxing and others},
  journal={arXiv preprint arXiv:2602.15763},
  year={2026}
}

@misc{minimax2026m27,
  author       = {{MiniMax}},
  title        = {{MiniMax M2.7}: Early Echoes of Self-Evolution},
  year         = {2026},
  month        = mar,
  howpublished = {\url{https://www.minimax.io/news/minimax-m27-en}},
  note         = {Accessed: 2026-10-01}
}

@article{team2026kimi,
  title={Kimi k2. 5: Visual agentic intelligence},
  author={Team, Kimi and Bai, Tongtong and Bai, Yifan and Bao, Yiping and Cai, SH and Cao, Yuan and Chai, Ziwei and Charles, Y and Che, HS and Chen, Cheng and others},
  journal={arXiv preprint arXiv:2602.02276},
  year={2026}
}

@misc{wang2025opencuaopenfoundationscomputeruse,
      title={OpenCUA: Open Foundations for Computer-Use Agents}, 
      author={Xinyuan Wang and Bowen Wang and Dunjie Lu and Junlin Yang and Tianbao Xie and Junli Wang and Jiaqi Deng and Xiaole Guo and Yiheng Xu and Chen Henry Wu and Zhennan Shen and Zhuokai Li and Ryan Li and Xiaochuan Li and Junda Chen and Boyuan Zheng and Peihang Li and Fangyu Lei and Ruisheng Cao and Yeqiao Fu and Dongchan Shin and Martin Shin and Jiarui Hu and Yuyan Wang and Jixuan Chen and Yuxiao Ye and Danyang Zhang and Dikang Du and Hao Hu and Huarong Chen and Zaida Zhou and Haotian Yao and Ziwei Chen and Qizheng Gu and Yipu Wang and Heng Wang and Diyi Yang and Victor Zhong and Flood Sung and Y. Charles and Zhilin Yang and Tao Yu},
      year={2025},
      eprint={2508.09123},
      archivePrefix={arXiv},
      primaryClass={cs.AI},
      url={https://arxiv.org/abs/2508.09123}, 
}

@article{xuan2014towards,
  title={Towards effective bug triage with software data reduction techniques},
  author={Xuan, Jifeng and Jiang, He and Hu, Yan and Ren, Zhilei and Zou, Weiqin and Luo, Zhongxuan and Wu, Xindong},
  journal={IEEE transactions on knowledge and data engineering},
  volume={27},
  number={1},
  pages={264--280},
  year={2014},
  publisher={IEEE}
}

@inproceedings{laiq2022early,
  title={Early identification of invalid bug reports in industrial settings--a case study},
  author={Laiq, Muhammad and Ali, Nauman bin and B{\"o}stler, J{\"u}rgen and Engstr{\"o}m, Emelie},
  booktitle={International Conference on Product-Focused Software Process Improvement},
  pages={497--507},
  year={2022},
  organization={Springer}
}

@article{zhang2026code,
  title={Code review agent benchmark},
  author={Zhang, Yuntong and Pan, Zhiyuan and Yusuf, Imam Nur Bani and Ruan, Haifeng and Shariffdeen, Ridwan and Roychoudhury, Abhik},
  journal={arXiv preprint arXiv:2603.23448},
  year={2026}
}

@inproceedings{shi2025judging,
  title={Judging the judges: A systematic study of position bias in llm-as-a-judge},
  author={Shi, Lin and Ma, Chiyu and Liang, Wenhua and Diao, Xingjian and Ma, Weicheng and Vosoughi, Soroush},
  booktitle={Proceedings of the 14th International Joint Conference on Natural Language Processing and the 4th Conference of the Asia-Pacific Chapter of the Association for Computational Linguistics},
  pages={292--314},
  year={2025}
}

@article{guo2026swe,
  title={SWE-Doctor: Guiding Software Engineering Agents with Runtime Diagnosis from Multi-Faceted Bug Reproduction Tests},
  author={Guo, Yaoqi and Liu, Yang and Zhang, Jie M and Ma, Yun and Lou, Yiling and Chen, Zhenpeng},
  journal={arXiv preprint arXiv:2607.00990},
  year={2026}
}

@article{cheng2025agentic,
  title={Agentic bug reproduction for effective automated program repair at google},
  author={Cheng, Runxiang and Tufano, Michele and Cito, J{\"u}rgen and Cambronero, Jos{\'e} and Rondon, Pat and Wei, Renyao and Sun, Aaron and Chandra, Satish},
  journal={arXiv preprint arXiv:2502.01821},
  year={2025}
}

@article{yadavally2025large,
  title={Large language model critics for execution-free evaluation of code changes},
  author={Yadavally, Aashish and Nguyen, Hoan and Callot, Laurent and Guinet, Gauthier},
  journal={arXiv preprint arXiv:2501.16655},
  year={2025}
}

@inproceedings{erfani2014works,
  title={Works for me! characterizing non-reproducible bug reports},
  author={Erfani Joorabchi, Mona and Mirzaaghaei, Mehdi and Mesbah, Ali},
  booktitle={Proceedings of the 11th working conference on mining software repositories},
  pages={62--71},
  year={2014}
}

@inproceedings{aktas2023issue,
  title={Issue report validation in an industrial context},
  author={Aktas, Ethem Utku and Cakmak, Ebru and Inan, Mete Cihad and Yilmaz, Cemal},
  booktitle={Proceedings of the 31st ACM joint european software engineering conference and symposium on the foundations of software engineering},
  pages={2026--2031},
  year={2023}
}

@article{zhang2026swe,
  title={Swe-bench goes live!},
  author={Zhang, Linghao and He, Shilin and Zhang, Chaoyun and Kang, Yu and Li, Bowen and Xie, Chengxing and Wang, Junhao and Wang, Maoquan and Huang, Yufan and Fu, Shengyu and others},
  journal={Advances in Neural Information Processing Systems},
  volume={38},
  year={2026}
}

@article{bruno2026writing,
  title={Writing Bug Reports for Software Repair Agents: What Information Matters Most?},
  author={Bruno, Vincenzo Luigi and Giagnorio, Alessandro and Bifolco, Daniele and Wienges, Leon and Di Penta, Massimiliano and Bavota, Gabriele},
  journal={arXiv preprint arXiv:2607.09553},
  year={2026}
}

@article{zhang2019bertscore,
  title={Bertscore: Evaluating text generation with bert},
  author={Zhang, Tianyi and Kishore, Varsha and Wu, Felix and Weinberger, Kilian Q and Artzi, Yoav},
  journal={arXiv preprint arXiv:1904.09675},
  year={2019}
}

@misc{stackoverflow2025survey,
  title        = {Stack Overflow Developer Survey 2025: AI},
  author       = {{Stack Overflow}},
  year         = {2025},
  howpublished = {\url{https://survey.stackoverflow.co/2025/ai}},
  note         = {Accessed: 2026-09-08}
}

@article{cohen1960coefficient,
  title={A coefficient of agreement for nominal scales},
  author={Cohen, Jacob},
  journal={Educational and psychological measurement},
  volume={20},
  number={1},
  pages={37--46},
  year={1960},
  publisher={Sage Publications Sage CA: Thousand Oaks, CA}
}

@article{wilson1927probable,
  title={Probable inference, the law of succession, and statistical inference},
  author={Wilson, Edwin B},
  journal={Journal of the American Statistical Association},
  volume={22},
  number={158},
  pages={209--212},
  year={1927},
  publisher={Taylor \& Francis}
}

\end{document}